\documentclass[conference]{IEEEtran}

\usepackage{cite}
\usepackage{graphicx}    %\usepackage[pdftex]{graphicx}
\usepackage{url}
\usepackage[intlimits]{amsmath}
\usepackage{color}
\usepackage{amssymb}
\begin{document}

\title{Grid-based Network Architecture for Distributed Computation %of Primitive Recursive Functions and its Application
in Wireless Sensor Networks}

\author{
\IEEEauthorblockN{
Rama Murthy Garimella,
Deepti Singhal
}
\IEEEauthorblockA{
International Institute of Information Technology, Hyderabad, India - 500 032.\\
\emph{rammurthy@iiit.ac.in}
}
}

% make the title area
\maketitle

\begin{abstract}
Wireless Sensor Networks (WSNs) are used to perform distributed sensing in various fields, such as health, military, home etc.  %Sensing is done in order to have a better understanding of the behavior of the monitored entity or to monitor an environment for the occurrence of a set of possible events, so that the proper action may be taken whenever necessary. Large amount of sensors are deployed in the sensor field to monitor the specified event, and to collectively take a decision, these sensor nodes should communicate and do distributed computation over the sensed values.
%For most of the monitoring situations, such as identification of fire or snowfall, the maximum/minimum of the monitored values should be required.
In WSNs, sensor nodes should communicate among themselves and do distributed computation over the sensed values to identify the occurrence of an event. This paper assumes the no memory computation model for sensor nodes, i.e. the sensor nodes only have two registers. This paper presents an optimal architecture for the distributed computation in WSN and also claims that this architecture is the optimal for the described computation model.\\

\end{abstract}

\IEEEpeerreviewmaketitle

\section{Introduction} \label{sec:intro}
WSN consists of a set of sensor nodes that are deployed in a field and interconnected with a wireless communication network. Each of these scattered sensor node has the capabilities to collect data, fuse that data and route the data back to the sink/base station \cite{1,2}. To collect data, each of these sensor nodes makes decision based on its observation of a part of the environment and on partial a-priori information. As larger amount of sensors are deployed in harsher environment, it is important that the distributed computation should be robust and fault-tolerant.
%In this paper we consider maximum and minimum of sensed values as a distributed computation function.
The identification of event in wireless sensor network should be done as fast as possible thus the computations are done in parallel.\\

Here we investigate the problem of design of optimal parallel distributed computation architecture. %for fire/snowfall detection.
In distributed system components located on networked computers communicate and coordinate by passing messages to perform the specified task. Similarly distributed computation is done on distributed nodes connected over the network with defined computation model. A model of computation \cite{computability} is a formal description of a particular type of computational process. This paper assumes the no memory computation model for sensor nodes. No memory computation model means the sensor node just have registers to store two values, whenever sensor node receives any value from other sensor node it simply compute the function with its own measured value and the received value and pass the results to other sensor node(s). \\

The distributed architecture for WSN needs to be optimal from most of the following points \cite{4}:
\begin{itemize}
    \item Computation complexity %Energy Efficiency
    \item Transmission delay required for the computations
    \item Deployment / Reconfiguration
    \item Fault Tolerance
\end{itemize}
%$~$ \\
%Computation complexity, measured in terms of number of comparisons required, can be used to measure energy efficiency.\\

The rest of the paper is a follows: Section \ref{sec:archi} describes the problem statement and discusses the optimal architecture. Section \ref{sec:PRF} discusses the class of functions, i.e. primitive recursive functions, which can be solved using grid like architecture.
%Section \ref{sec:PM} discusses the tradeoff between the performance measures of distributed computation mainly computation complexity and transmission delay.
Fault tolerance capability of the proposed architecture is discussed in section \ref{sec:fault}. %Finally section \ref{sec:future} \& \ref{sec:conclusion} presents the future scope of the work and conclusion of the paper.
Finally section \ref{sec:conclusion} concludes the paper.

%\section{Tradeoff between the Performance Measures} \label{sec:PM}
%This section discusses the tradeoff between the transmission delay and the computation delay. Let us take an example to understand the tradeoff between these two performance measures. The problem is to compute the Maximum/Minimum of $N$ numbers distributed over $N$ processors. Assuming two scenarios, first the processors are connected in line and second the processors are connected using binary tree architecture as shown in figure \ref{fig:topo}.
%\begin{figure}[h!]
%    \centerline {
%    \scalebox {0.5} {
%        \includegraphics{topo.eps}
%    }
%    }
%    \caption{Different topology of distributed processors}
%    \label{fig:topo}
%\end{figure}
%Assume that the delay in each link is $d$. The calculations for both the topologies are as follows:
%\begin{itemize}
%  \item The Transmission delay in line topology is $(N-1) d$ while in tree type topology it is $(log_2N) d$.
%  \item The Computation Complexity in line topology is $(N-1)$ while in tree type topology it is $N log_2N$.
%\end{itemize}
%$~$\\
%The transmission delay in binary tree topology decreases as in every sub-tree, computations can be done in parallel. But the computation complexity increases as for calculating maximum/minimum sorting should be performed. The second topology is studied under distributed sorting problems \cite{4,5}.\\
%
%Here we can see decreasing the transmission delay will increase the computation complexity.\\

\section{Distributed Network Architecture} \label{sec:archi}
\subsection{Problem Statement}
The problem is to define a globally optimal data structure for calculating the defined fusion function over the sensor field. The architecture should be as optimal as possible from the point of view of all the performance measures discussed in section \ref{sec:intro}.
%As we have seen in section \ref{sec:PM} that computation complexity and the transmission delay has the tradeoff. In such tradeoff situations, to find the globally optimal architecture, we need to fix one and try to optimize the other parameter. 
The computation model considered is also important while defining the suitable distributed architecture. The paper assumes the no memory computation model as discussed in section \ref{sec:intro}. Thus we are redefining the problem statement. %by fixing the maximum allowed delay. 
To find the globally optimal architecture, we need to fix some of the performance measures and try to optimize the other measures. The modified problem statement is:
\begin{quote}
    Given the maximum allowed delay $D_0$, define the globally optimum data structure of wireless sensor network for the distributed computation of fusion functions %discussed in section \ref{sec:PRF} 
    of sensed values in no memory computation model. % of interest in fire control WSN or snowfall detection WSN.
\end{quote}

\subsection{Grid based Network Architecture} \label{subsec:archi}
This sections discusses the optimal distributed architecture for homogeneous and heterogeneous wireless sensor networks. Homogeneous WSN consists of sensor nodes with same ability while heterogeneous WSN consists of sensor nodes with different ability such as different computing power.

\subsubsection{Homogeneous WSN} 
\begin{figure}[h]
    \centerline {
    \scalebox {0.5} {
        \includegraphics{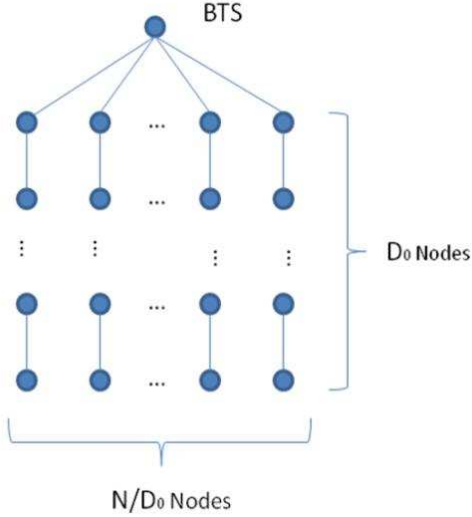}
    }
    }
    \caption{Grid Based Architecture}
    \label{fig:archi}
\end{figure}

The solution for the above defined problem is a grid like architecture as shown in figure \ref{fig:archi}.

In this section, we first discuss the computation complexity of the minimum/maximum like fusion function where only one comparison is need at every node, later we show that the same comparisons can be done for other %primitive recursive
functions. Assume that the total number of processors $P$ is equal to the number of sensor nodes $N$ in the network. Calculations are as follows:\\
The number of nodes in each branch $= D_0$\\
Computation complexity in each branch is $= (D_0 - 1)$\\
Total number of such branches $= \frac{N}{D_0}$\\
So, total computation complexity\\
\begin{equation*}
    = \frac{N}{D_0}(D_0 - 1) + \frac{N}{D_0 -1} %comparisons
\end{equation*}
\begin{equation*}
= N - \frac{N}{D_0} + \frac{N}{D_0} - 1 %comparisons\\
\end{equation*}
\begin{equation*}
= N - 1 %comparisons\\
\end{equation*}
$~$\\
%So, total computation complexity\\
%$= \frac{N}{D_0}*(D_0 - 1) + \frac{N}{D_0 -1}$ comparisons\\
%$= N - \frac{N}{D_0} + \frac{N}{D_0} - 1$ comparisons\\
%$= N - 1$ comparisons\\
%Here we can see that the number of comparisons is equal to the comparisons required for the line architecture which is the lower bound of the computation complexity. Also the architecture is maintaining the transmission delay requirement to $D_0$.\\
We can see that the number of comparisons is equal to the minimum comparisons required for any architecture which is the lower bound of the computation complexity. Also the architecture is maintaining the transmission delay requirement to $D_0$. This is also possible by tree kind of architecture. In tree architecture for maintaining the delay requirement, one node will have multiple child nodes. And as any sensor node will receive more number of values simultaneously, more number of registers are needed and hence tree kind of architectures are not suitable in this computation model. \\

Now let us consider the special class of function which require $x$ comparisons at each sensor node. Calculation for the computation complexity in such functions are: \\
The number of nodes in each branch $= D_0$\\
Computation complexity in each branch is $= (D_0 - 1)*x$\\
Total number of such branches $= \frac{N}{D_0}$\\
So, total computation complexity\\
\begin{equation*}
    = \frac{N}{D_0}(D_0 - 1)x + \frac{N}{D_0 -1}x %comparisons
\end{equation*}
\begin{equation*}
= Nx - \frac{N}{D_0}x + \frac{N}{D_0}x - x %comparisons\\
\end{equation*}
\begin{equation*}
= (N - 1)x %comparisons\\
\end{equation*}
$~$\\
Here again the number of comparisons is equal to the minimum required comparisons.% required for the line architecture which is the lower bound of the computation complexity. 
Also this architecture is very suitable for the sensor field from the point of view of deployment and coverage.\\

\subsubsection{Heterogeneous WSN}
Clustering of sensor networks has proved to be very effective in conserving energy in heterogeneous networks. For each cluster, one Cluster Head(CH) is selected using \cite{leach,head}. CH is responsible for collecting, fusing and transmission of data for the cluster, and also only cluster heads participate in routing the other cluster data to the base station/sink. Thus in heterogeneous WSN, we consider three type of nodes:
\begin{enumerate}
  \item Sensor Nodes
  \item Cluster Heads
  \item Base Station/Sink
\end{enumerate}

\begin{figure}[h!]
    \centerline {
    \scalebox {0.45} {
        \includegraphics{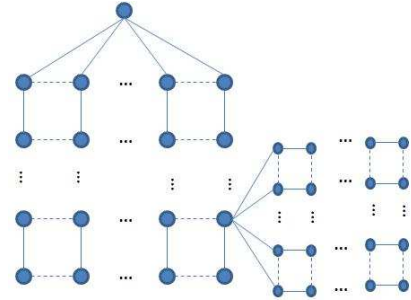}
    }
    }
    \caption{Grid Architecture for Heterogeneous WSNs}
    \label{fig:archi_hetero}
\end{figure}

The proposed architecture for this kind of network consists of hierarchy of grid architecture as shown in figure \ref{fig:archi_hetero}. In the hierarchical grid architecture, each node of the final grid architecture is actually a cluster head and this cluster head is connected to other sensor nodes of the cluster using other grid based architecture. \\

Similarly for more complex networks, multiple hierarchy of clusters can be done and in such cases multi-hierarchical grid based architecture can be used.

%\subsection{Comparison with Balanced Binary Tree Architecture}
%Considering balanced binary tree of total N nodes in place of grid architecture as shown in figure \ref{fig:bbt}, the calculation is as follows:\\
%$N = 2(h+1) - 1$ where $h$ is the depth of the tree.\\
%$h = floor (log_2 N)$.\\
%Maximum delay bound achievable in balanced binary tree of total $N$ nodes is $= h = floor (log_2 N)$.  If $D_0 < h$, than binary tree architecture can't be used. But in grid architecture there is no limit on maximum delay bound achievable.
%
%\begin{figure}[h]
%    \centerline {
%    \scalebox {0.5} {
%        \includegraphics{bbt.eps}
%    }
%    }
%    \caption{Balanced Binary Tree Architecture}
%    \label{fig:bbt}
%\end{figure}
%
%Let us assume $D_0 = h = floor (log_2 N)$ then also complexity of finding maximum/minimum with balanced binary tree is $O(Nlog_2N)$ while for grid architecture it is $O(N - 1)$.\\
%
%Tree kind of architecture is also not suitable for the deployment in WSN as it is not possible to cover the whole area while grid kind of architecture is very suitable for the deployment in WSN.\\

\section{Primitive Recursive Functions } \label{sec:PRF}
This section discusses the class of functions, i.e. primitive recursive functions, which can be computed optimally on grid like architecture.
The basic primitive recursive functions are given by these axioms \cite{10}:
\begin{itemize}
    \item \textbf{Constant function:} The $0-ary$ constant function $0$ is primitive recursive.
    \item \textbf{Successor function:} The $1-ary$ successor function $S$, which returns the successor of its argument, is primitive recursive. That is, $S(k) = k + 1$.
    \item \textbf{Projection function:} For every $n\geq 1$ and each $i$ with $1 \leq i \leq n$, the $n-ary$ projection function $(P_i)^n$, which returns its $i^(th)$ argument, is primitive recursive.
\end{itemize}

More complex primitive recursive functions can be obtained from the initial functions by means of composition and primitive recursion.
\begin{itemize}
  \item \textbf{Composition:} If $g$ is a function of $m$ arguments, and each of $h_1,…,h_m$ is a function of $n$ arguments, then the function $f$ $f(x_1,…,x_n) = g(h_1(x_1,…,x_n), … , h_m(x_1,…,x_n))$
is definable by composition from $g$ and $h_1,…,h_m$. We write $f = [g \circ h_1,…,h_m]$, and in the simple case where $m = 1$ and $h_1$ is designated $h$, we write $f(x) = [g \circ h](x)$
  \item \textbf{Primitive Recursion:} A function $f$ is definable by primitive recursion from $g$ and $h$ if:
  \begin{eqnarray}
  % \nonumber to remove numbering (before each equation)
    \nonumber f(x,0)  &=& g(x) \\
    \nonumber f(x,s(y)) &=& h(x, y, f(x,y))
  \end{eqnarray}
We write $f = PR[g, h]$ when $f$ is definable by primitive recursion from $g$ and $h$. Here $s$ is the successor function, which when given an argument $n$ returns its immediate successor $s(n)$.
\end{itemize}
$~$\\
The primitive recursive functions are the basic functions and those obtained from the basic functions by applying these operations a finite number of times. In WSN domain, simple aggregation techniques i.e., maximum, minimum, and average have been used to save energy while monitoring \cite{WSN_Fusion}. Incase of more complex fusion functions also, the fusion function can be represented using primitive recursive function class. Some of the examples of primitive recursive functions which can be used in fusion are:
\begin{itemize}
  \item Addition
  \item Multiplication
  \item Exponentiation
  \item Factorial
  \item Proper subtraction defined as `If $a \geq b$ then $a-b$ else $0$.'
  \item Minimum $(a_1, \ldots, a_n)$
  \item Maximum $(a_1, \ldots, a_n)$
  \item Absolute value
  \item Mean
  \item Weighted Mean
  \item Weighted Energy, i.e. $w_1 x_1^2 + w_2 x_2^2 + \ldots + w_m x_m^2$
\end{itemize}
$~$\\

\section{Fault Tolerance of Proposed Architecture} \label{sec:fault}
%
%Failures are not an exception in WSN. For example, consider a WSN that monitors
%a jungle to detect an event, such as fire or the presence of an animal. Sensor nodes
%can be destroyed by fire, animals, or even human beings; they might present
%manufacturing problems; and they might stop working due to a lack of energy.

\begin{figure}[h]
    \centerline {
    \scalebox {0.5} {
        \includegraphics{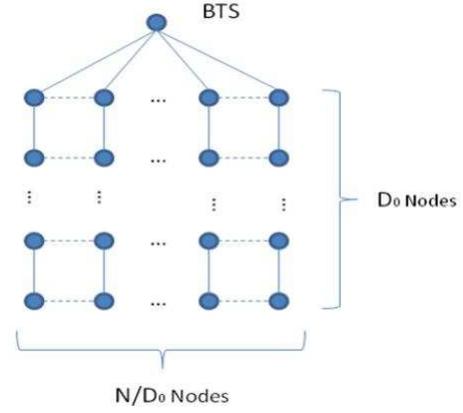}
    }
    }
    \caption{Fault Tolerant Grid Architecture}
    \label{fig:archi2}
\end{figure}

Fault tolerance of a network is a measure of its performance to do the intended job if some node(s), link(s) or both fail. To increase the fault tolerance of the proposed architecture we included the row links as shown in figure \ref{fig:archi2}. In this architecture if a node or link goes down, then other path is available to send the value computed so far which can be used while calculating the fusion function of the other branch.\\

Also we can do identification of error in this architecture. The steps for error detection are as follows:
\begin{enumerate}
  \item Calculate row wise fusion function
  \begin{itemize}
    \item Calculate the row wise fusion function for each row.
    \item Calculate the fusion function with each row's calculated value.
  \end{itemize}
  \item Similarly Calculate column wise fusion function
  \begin{itemize}
    \item Calculate the column wise fusion function for each column.
    \item Calculate the fusion function with each column's calculated value.
  \end{itemize}
  \item If the result of step 1 and 2 are not same than there is some error in the computation.
\end{enumerate}
$~$\\

\subsection{Example of fault tolerance}
\begin{figure}[h]
    \centerline {
    \scalebox {0.4} {
        \includegraphics{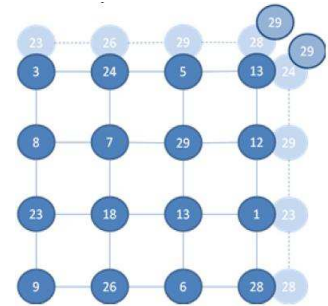}
    }
    }
    \caption{Fault Tolerance Example}
    \label{fig:ft1}
\end{figure}

%Let us take an example to understand the fault tolerance of the architecture as shown in figure \ref{fig:ft1}. This figure shows the correct functionality without any link or node failure.
This example consider the `maximum' as the fusion function to be computed. Figure \ref{fig:ft1} shows the correct functionality without any link or node failure.

\begin{figure}[h!]
    \centerline {
    \scalebox {0.8} {
        \includegraphics{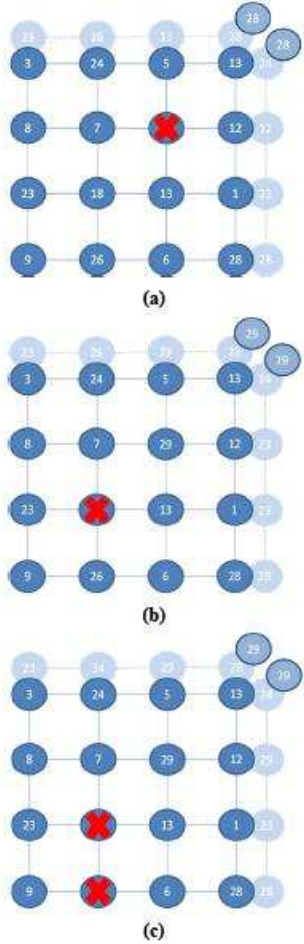}
    }
    }
    \caption{Fault Tolerance Example: Node Failure Scenario}
    \label{fig:ft2}
\end{figure}

Figure \ref{fig:ft2} shows the node failure scenario of the example. Node failure is considered where the reading from a particular node is not available. The node failure doesn't provide the reading for the respective node. If the failure node is the one giving maximum value as shown in the figure \ref{fig:ft2}(a), than this affects the distributed computation and can't be detected. But if some other node(s) (not giving the maximum value) fails than the computation is correct and hence the node failure can't be detected. It is not such a restrictive idea to assume that the maximum over the sensor field is sensed by a single node. But multiple nodes could also sense the maximum value. Here again if all the node giving maximum value fails than only the computation is incorrect otherwise the computation is correct.\\

\begin{figure}[h]
    \centerline {
    \scalebox {0.8} {
        \includegraphics{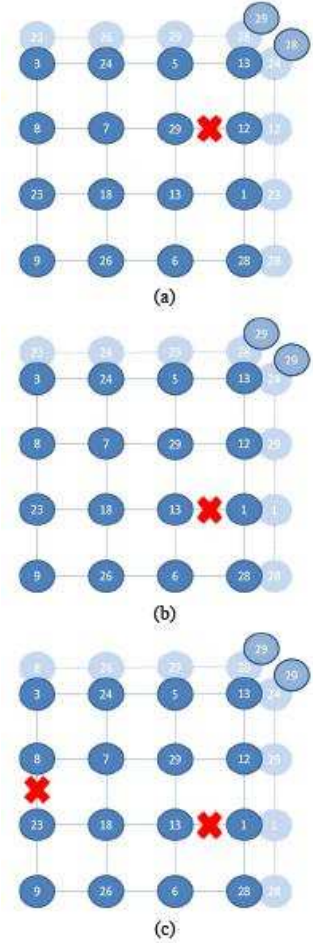}
    }
    }
    \caption{Fault Tolerance Example: Link Failure Scenario}
    \label{fig:ft3}
\end{figure}

Figure \ref{fig:ft3} shows the link failure scenario of the example. If the failure link is the one passing the maximum value as shown in the figure \ref{fig:ft3}(a), than this can be detected. But if some other link(s) (not passing the maximum value) fails than the computation is correct and hence the link failure can't be detected.\\

From this example it is clear that the one link failure (if the one is passing the maximum value) can be detected in the architecture.\\

\subsection{Probability error in maximum calculation}
This subsection calculate the probability of error in grid based architecture for maximum calculation.\\
Total number of sensor nodes $=N$ \\
Lets assume nodes fail independently with probability $= P_f$ \\
And the number of nodes containing the maximum value $= M$ \\
So the probability of error is :
\begin{equation*}\label{err}
   P_e =\frac{1}{^{N}C_{M}} p_f^M (1 - p_f)^{(N - M)}
\end{equation*}
\begin{equation*}\label{err}
   P_e =\frac{M! (N-M)!}{N!} p_f^M (1 - p_f)^{(N - M)}
\end{equation*}
Here if only one node contain the maximum value, then the probability of error is $\frac{1}{N} p_f (1 - p_f)^{(N - 1)}$. In grid the probability of no error is $(1 - P_e)$. \\
%\section{Scope of Future Work} \label{sec:future}
%The future research work is carried out on localization of error, i.e. the identification of node(s) or link(s) which is not working as intended. One idea to do this by probing the grid architecture with test data and testing it, this will give some probabilistic measures for each link and node failures.\\

\section{Conclusion}\label{sec:conclusion}
This paper discusses the architecture for distributed computation %of Maximum/Minimum function 
in wireless sensor network. The result for computation complexity for fixed maximum allowed delay is calculated in section \ref{subsec:archi}, which shows that this architecture is the solution for given computation model. Section \ref{sec:fault} shows that the proposed architecture is very efficient with respect to faults in the network.\\

%The problem of determining the maximum/minimum of a list of numbers (e.g. sensed values) arises in many real world scenarios. Our grid based architecture is a natural choice.\\

%\section*{Acknowledgment}

\bibliographystyle{unsrt} %plain}
\bibliography{biblist}

\begin{thebibliography}{1}

\bibitem{1}
I.F. Akyildiz, Weilian Su, Y.~Sankarasubramaniam, and E.~Cayirci.
\newblock A survey on sensor networks.
\newblock {\em Communications Magazine, IEEE}, 40(8):102--114, 2002.

\bibitem{2}
I.~F. Akyildiz, W.~Su, Y.~Sankarasubramaniam, and E.~Cayirci.
\newblock Wireless sensor networks: a survey.
\newblock {\em Computer Networks}, 38:393--422, 2002.

\bibitem{computability}
S.B. Cooper.
\newblock {\em Computability Theory}.
\newblock Chapman \& Hall/CRC mathematics. Chapman \& Hall/CRC, 2004.

\bibitem{4}
D.~Rotem, N.~Santoro, and Jeffrey~B. Sidney.
\newblock Distributed sorting.
\newblock {\em Computers, IEEE Transactions on}, C-34(4):372--376, 1985.

\bibitem{leach}
W.R. Heinzelman, A.~Chandrakasan, and H.~Balakrishnan.
\newblock Energy-efficient communication protocol for wireless microsensor
  networks.
\newblock In {\em System Sciences, 2000. Proceedings of the 33rd Annual Hawaii
  International Conference on}, pages 10 pp. vol.2--, 2000.

\bibitem{head}
O.~Younis and Sonia Fahmy.
\newblock Heed: a hybrid, energy-efficient, distributed clustering approach for
  ad hoc sensor networks.
\newblock {\em Mobile Computing, IEEE Transactions on}, 3(4):366--379, 2004.

\bibitem{10}
Piergiorgio Odifreddi and S.~Barry Cooper.
\newblock Recursive functions.
\newblock In Edward~N. Zalta, editor, {\em The Stanford Encyclopedia of
  Philosophy}. Fall 2012 edition, 2012.

\bibitem{WSN_Fusion}
Ahmed Abdelgawad and Magdy Bayoumi.
\newblock Data fusion in wsn.
\newblock In {\em Resource-Aware Data Fusion Algorithms for Wireless Sensor
  Networks}, volume 118 of {\em Lecture Notes in Electrical Engineering}, pages
  17--35. Springer US, 2012.

\end{thebibliography}

\end{document}